# EFFECT OF SURFACE FINISH OF SUBSTRATE ON MECHANICAL RELIABILITY OF IN-48SN SOLDER JOINTS IN MOEMS PACKAGE

*Ja-Myeong Koo and Seung-Boo Jung*

School of Advanced Materials Science and Engineering, Sungkyunkwan University,
300 Cheoncheon-dong, Jangan-gu, Suwon 440-746, Republic of Korea

**ABSTRACT**

Interfacial reactions and shear properties of the In-48Sn (in wt.%) ball grid array (BGA) solder joints after bonding were investigated with four different surface finishes of the substrate over an underlying Cu pad: electroplated Ni/Au (hereafter E-NG), electroless Ni/immersion Au (hereafter ENIG), immersion Ag (hereafter I-Ag) and organic solderability preservative (hereafter OSP). During bonding, continuous $AuIn_2$, $Ni_3(Sn,In)_4$ and $Cu_6(Sn,In)_5$ intermetallic compound (IMC) layers were formed at the solder/E-NG, solder/ENIG and solder/OSP interface, respectively. The interfacial reactions between the solder and I-Ag substrate during bonding resulted in the formation of $Cu_6(Sn,In)_5$ and $Cu(Sn,In)_2$ IMCs with a minor Ag element. The In-48Sn/I-Ag solder joint showed the best shear properties among the four solder joints after bonding, whereas the solder/ENIG solder joint exhibited the weakest mechanical integrity.

## 1. INTRODUCTION

The growth of fiber optic networks is driving the development of novel miniaturized and high performance optical components based on micro electro mechanical system (MEMS) technology [1]. These needs for the micro optical electro mechanical systems (MOEMS) have pushed the development of area-array packages, such as ball grid array (BGA), wafer level package (WLP) and flip-chip, due to their smaller package size, more input/output (I/O) pins and higher electrical performance. For the MOEM package, the area array package using solder materials is simpler to process than the thermosonic bonding process using gold or aluminum metal bumps, and has better electrical performance and mechanical properties than the bonding process using conductive adhesives. In the area-array package, the solder joints provide a path for the dissipation of the heat generated by the device as well as furnish the electrical and mechanical connection between the device and substrate [2]. Therefore, the reliability of solder joints has been a crucial issue in the MEMS and MOEMS packaging industries.

Essential solder materials for the MOEM package can be categorized as tin-lead (Pb-Sn), gold-tin (Au-Sn), indium (In) and indium-tin (In-Sn) [3]. The solder assembly using Sn-Pb solder has been the most popular approach. However, electrical and electronic equipment (EEE) manufacturers are showing more and more interest in the development of Pb-free solders and their soldering processes, due to the increasing environmental awareness of society, upcoming legislation on the use of Pb such as the waste of electrical and electronic equipment (WEEE) and restriction of hazardous substances (RoHS) directives and the tremendous market potential for 'green' products [4]. The Au-Sn solder shows low-creep behavior, but induces the residual stress at the joint interface during bonding and system use due to its high stiffness and bonding temperature. Pure In solder exhibits excellent heat and electrical conductivity, but has poor creep resistance [3]. In this respect, eutectic In-Sn (In-48 wt.% Sn) solder alloy has been considered as a Pb-free solder material with good potential in the MOEMS package, due to its very low melting point, great ductility and long fatigue life [3-5]. While industrial interest in the In-Sn solder is increasing, there has been relatively little fundamental research on its bonding characteristics and mechanical reliability [5].

During the bonding process, it is inevitable that the intermetallic compounds (IMCs) form and grow at the solder/substrate interface [6]. The formation of a thin IMC layer between the molten solder and substrate during bonding is essential to the bondability of these two materials, while the excessive IMC formation at the interface weakens the solder joints. The metallurgical and mechanical properties of the solder joints formed during bonding significantly depend on the surface finish of the substrate as well as the solder material. With the adoption of Pb-free solders in the packaging technology, the selection of an appropriate metallization on a substrate plays an increasingly important role in the development of reliable solder joints [7]. Because the In-48Sn (in wt.%) alloy has the low solubility and dissolution rate of Au in the molten solder at a bonding temperature, the Au layer, which is a typical surface finish on the substrate in the MOEMS package, reacted with the In-48Sn solder to form a brittle $AuIn_2$ IMC layer at the interface [4]. Therefore, the application of the In-48Sn solder joint has been limited in MEMS and MOEMS systems. The OSP surface finish has several advantages, such as its good wetting property, low cost and simple processing steps, while the In-Sn/Cu system exhibited fast IMC growth to consume the Cu layer rapidly [8]. The In-Sn/ENIG solder joint may provide the key to successful application of In-48Sn solder, due to its thin Au layer, while the $Ni_3P$ formed at the interface between $Ni_3(Sn,In)_4$ IMC and Ni-P substrate can weaken the solder joint [9]. The I-Ag surface finish provides excellent solderability and electrical properties, but neither the interfacial reaction nor mechanical property of In-Sn/I-Ag solder joint has been reported yet.

The purpose of this work is to investigate the interfacial reactions and shear properties of the In-48Sn solder joints with the surface finish on the BGA substrate during bonding, for the application of the In-48Sn solder in MOEM packages. Four commercial surface finishes over an underlying Cu pad were examined in this work: E-NG, ENIG, I-Ag and OSP.

## 2. EXPERIMENTAL PROCEDURE

### 2.1. Sample Preparation

The solder balls used in this study were In-48Sn (in wt.%) solder spheres having a diameter of 500 μm (Indium Corporation, U.S.A.). The substrate was a FR-4 BGA substrate. The solder bonding pad was designed as a solder mask defined (SMD) type with a pad opening of 460 μm in diameter, a pad pitch of 2 mm in length and a solder mask (or side wall) of 10 μm in thickness. Four surface finishes of the substrate over an underlying Cu (30 μm) pad were examined in this study: electroplated Ni (7 μm)/Au (0.7 μm), electroless Ni-P (5 μm)/immersion Au (0.15 μm), immersion Ag and organic solderability preservative (OSP). The electroless plated Ni layer contained about 15 at.% P and had an amorphous structure. The BGA substrates were cleaned using an ultra-sonic cleaner, and then dried with hot air. The solder balls were dipped into an RMA (Rosin Mildly Activated) flux and placed on the pads of the BGA substrates. All of the samples were bonded in a reflow machine (SAT-5100 + profile temperature raise heater, Rhesca Co., Japan). The heating rate, bonding temperature and time were 1.5 °C/s, 150 ± 3 °C and 100 s, respectively. After the bonding process, all samples were air-cooled to room temperature. The soldering process was carried out in a nitrogen atmosphere to prevent oxidation of the samples. The oxygen concentration in the bonding furnace was measured using an oxygen analyzer (MAXO2, Maxtec[TM], U.S.A.) and kept below 1000 ppm during the bonding process. The samples were cleaned using an ultrasonic cleaner with a flux remover after the bonding process.

### 2.1. Observation of Microstructure

Upon completion of the bonding process, the specimens were mounted in cold epoxy, ground using 100, 400, 1200, 1500 and 2,000-grit SiC papers through a row of solder balls and polished with 0.3 μm $Al_2O_3$ powder. The microstructures of the samples were observed using scanning electron microscopy (SEM) in back-scattered electron imaging mode (BEI). For more accurate observation of the microstructures, solder was selectively etched with an etching solution consisting of 80 vol.% $H_2O$, 10 vol.% HF and 10 vol.% $H_2O_2$. Elemental analysis was also carried out using an energy dispersive X-ray spectroscopy (EDS). For each sample, the IMC area formed at the interface was measured using an image analyzer (UTHCSA Image tool). The measured area was then divided by the width of the image, in order to estimate the average IMC layer thickness.

### 2.3. Ball Shear Test

The ball shear tests were conducted using a bonding tester (PTR-1000, Rhesca Co., Japan) at 3600±300 s after bonding, due to the influence of room temperature aging on the solder alloys. The experimental procedure for the BGA ball shear tests followed the JEDEC standard (JESD22-B117). Before testing, the 5 kgf load cell used in this study was calibrated using a 1 kgf standard weight. The displacement rate and probe height were 200 μm/s and 50 μm, respectively. The displacement at break was measured when the shear force had decreased by 67% (2/3) of its maximum value. The work of break was obtained from measuring the area of the F-x (shear force–displacement) curve using the image analyzer software. The cross-sections of the fractured specimens were observed using SEM.

## 3. RESULTS AND DISCUSSION

### 3.1. Microstructure

In-48Sn solders and four different surface-finished BGA substrates were bonded successfully in a nitrogen atmosphere. Figure 1 shows the SEM micrographs of the In-48Sn BGA solder joints with different surface finishes after bonding. Based on the EDS analysis results, the bulk solder exhibited a lamellar structure consisting of β ($In_3Sn$) and γ ($InSn_4$) phases.

After bonding, there was a continuous facetted IMC layer, having a thickness of approximate 1.3 μm, between the solder and the topmost Au layer of the E-NG substrate, as shown in Fig. 1(a). The chemical composition of the IMC layer was $In_{66.4}Au_{33.6}$ (in at.%), corresponding to $AuIn_2$. Generally, the Au layer of the E-NG substrate dissolved into the molten Sn-based solder to form $AuSn_4$ IMC particles at a bonding temperature of 225 °C within 30 s [9]. However, the topmost Au layer remained even after bonding at 150 °C for 100 s, because of very low solubility and dissolution rate of Au in the molten In-48Sn solder at 150 °C [10].

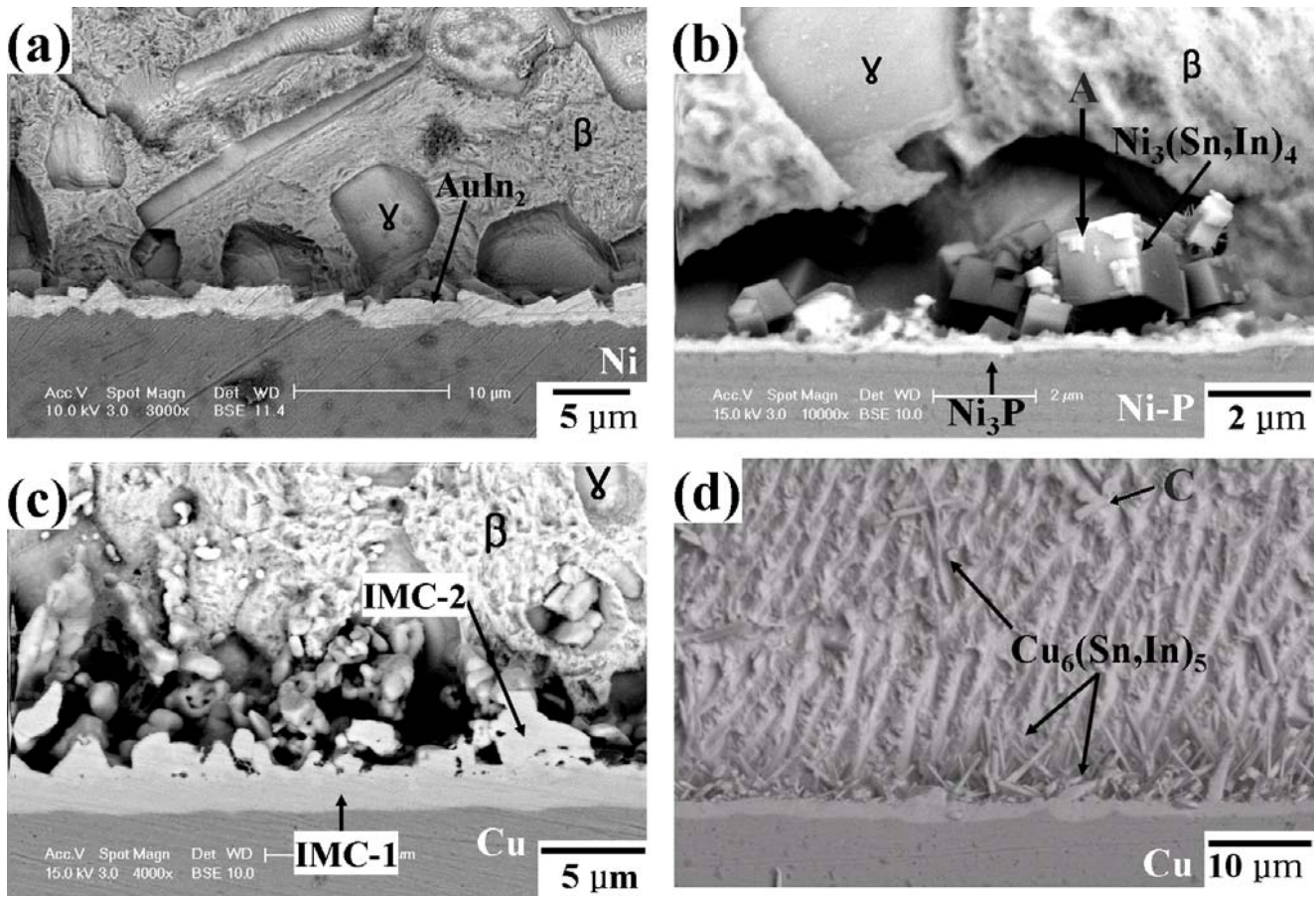

Figure 1 SEM micrographs of the interface between the In-48Sn (in wt.%) solder and BGA substrate with four different surface finishes over an underlying Cu pad after reflow: (a) E-NG, (b) ENIG, (c) I-Ag and (d) OSP.

All Au atoms of the ENIG substrate reacted with the molten solder and spalled off to the solder during bonding, due to the layer thinness. Therefore, neither Au nor Au-based IMC layer could be observed at the solder/substrate interface after bonding. The reaction between the molten solder and the exposed Ni layer resulted in the formation of a very thin IMC layer, having a thickness of approximate 0.21 µm, as shown Fig. 1(b). The chemical composition of the IMC layer consisted of $Sn_{45.8}Ni_{37.9}In_{16.3}$ (in at.%), corresponding to $Ni_3(Sn,In)_4$. Cubes of $Ni_3(Sn,In)_4$, marked as A in Fig. 1(b), existed over the thin $Ni_3(Sn,In)_4$ IMC layer. Generally, the diffusivity of atoms at the grain boundary was much faster than that of the same atoms in the lattice [11], suggesting that the cubes were formed by the Ni atoms diffusing through the grain boundaries of the IMC layer rapidly and subsequently reacted with the molten solder. A very thin reaction layer could be observed at the interface between the $Ni_3(Sn,In)_4$ and Ni-P layers. The reaction layer was too thin to determine its chemical composition exactly using EDS and EPMA analyses. It was reported in the previous study using transmission electron microscope (TEM) analysis that the reaction layer formed between the $Ni_3(Sn,In)_4$ and Ni-P layer during the liquid/solid reaction was $Ni_3P$ [12].

The topmost Ag layer of the I-Ag substrate was completely depleted and disappeared at the solder/substrate interface after bonding. It is suggested that the consumption of the Ag layer was faster than that of the Au layer during bonding. There were two different IMCs at the interface between the solder and the exposed Cu layer after bonding. Based on the EDS analysis results, the chemical compositions of the lower and upper IMC layers, marked as IMC-1 and IMC-2 in Fig. 1(c), respectively, consisted of $Cu_{58.1}Sn_{21.0}In_{15.2}Ag_{5.6}$ (in at.%), and $In_{44.1}Cu_{31.4}Sn_{16.6}Ag_{7.9}$ (in at.%). As the thicknesses of the IMC-1 and IMC-2 layers were approximate 1.7 μm and 1.2 μm, respectively, it is suggested that IMC-1 was the dominant phase in the In-48Sn/I-Ag interfacial reactions. These results were consistent with those in the previous studies. Kim et al. found that the solid-state reactions between the In-48Sn and Cu substrate resulted in the formation of $Cu(In,Sn)_2$ and $Cu_6(Sn,In)_5$ IMC layers, containing $In_{52}Cu_{32}Sn_{16}$ (in at.%) and $Cu_{56}Sn_{24}In_{20}$ (in at.%), respectively [13]. The chemical compositions of IMC-1 and IMC-2 in this work were similar to those of $Cu_6(Sn,In)_5$ and $Cu(In,Sn)_2$, respectively. This suggests that the IMC-1 and IMC-2 corresponded to $Cu_6(Sn,In)_5$ and $Cu(In,Sn)_2$, with a minor Ag element, respectively. Moreover, it was observed that the intermetallics, marked as B in Fig. 1(c), spalled off to the solder after bonding. The chemical compositions of the intermetallics were $Cu_{65.2}Sn_{13.3}In_{13.3}Ag_{8.2}$ (in at.%) and $In_{44.5}Ag_{24.7}Cu_{16.2}Sn_{14.6}$ (in at.%), respectively. The IMCs which spalled off to the solder during bonding had higher Ag content than the IMCs that existed at the solder/substrate interface.

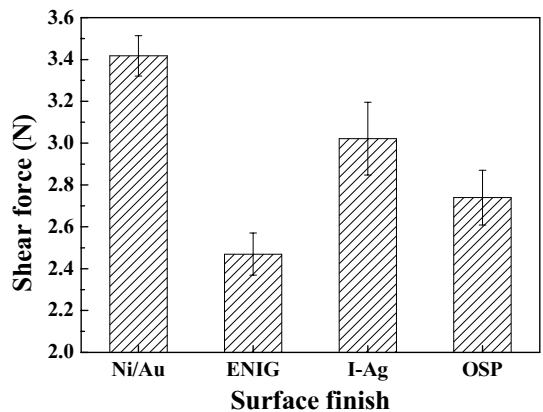

Figure 2 Shear force of the In-48Sn (in wt.%) solder joints with the surface finishes of the BGA substrate after bonding.

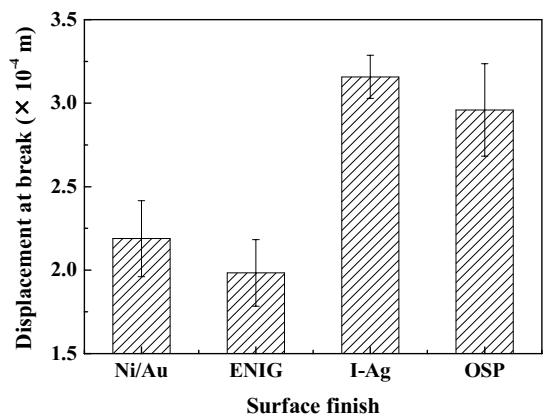

Figure 3 Displacement at break of the In-48Sn (in wt.%) solder joints with the surface finishes of the BGA substrate after bonding.

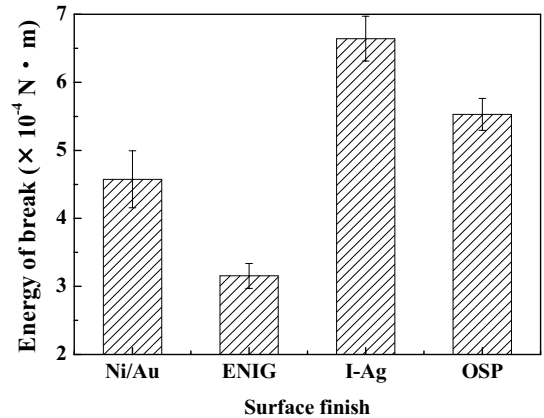

Figure 4 Energy of break of the In-48Sn (in wt.%) solder joints with the surface finishes of the BGA substrate after bonding.

The interfacial reaction of the In-48Sn solder with the OSP substrate resulted in the formation of two differently shaped IMCs, consisting of needle and planar types, as shown in Fig. 1(d). The chemical composition of the IMCs was $Cu_{57.4}Sn_{25.6}In_{17.0}$ (in at.%), corresponding to $Cu_6(Sn,In)_5$. Sommadossi et al. reported that only $Cu_6(Sn,In)_5$ IMC, having two different morphologies, was formed at the interface after bonding below 200 °C [14]. Laurila et al. suggested that the needle-shaped IMC was formed by the rapid dissolution of Cu in the molten solder followed by the local constitutional supercooling of the molten solder [15]. These results were consistent with our results. The needle-shaped $Cu_6(Sn,In)_5$ intermetallics, marked as C in Fig. 1(d), were also found in the bulk solder. It is suggested that these intermetallics spalled off to the molten solder during bonding.

### 3.2. Ball Shear Properties

The ball shear tests were carried out with four different surface finishes of the substrate, to investigate the effect of the surface finishes on the shear properties, such as shear force, displacement at break, work of break and fracture mode, of the In-48Sn BGA solder joints after bonding. The complex shape of the solder joint made it difficult to define the shear area and gage length. In this work, therefore, the shear force, displacement at break and work of break were used to evaluate the mechanical integrity of the solder joint, instead of the shear strength, elongation and toughness, respectively.

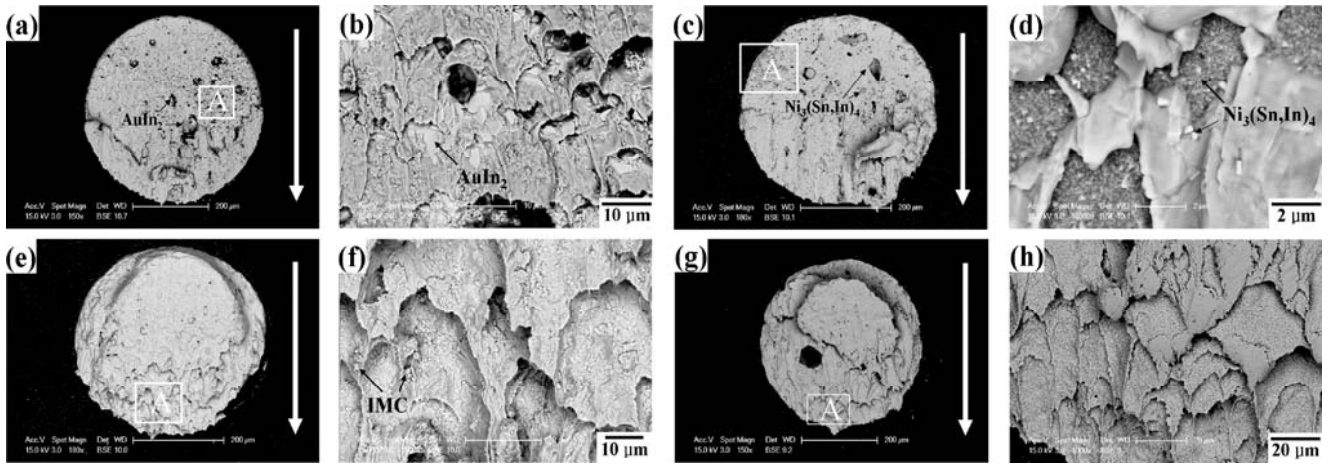

Figure 5 SEM micrographs of fracture surfaces of the shear-tested Sn-48Sn (in wt.%) BGA solder joints with four different surface finishes over an underlying Cu pad after bonding: (a) E-NG, (b) ENIG, (c) I-Ag and (d) OSP. Figures (b), (d), (f) and (h) show the magnified images of region A in Figs. (a), (c), (e) and (g), respectively.

Figure 2 shows the relationship between the shear force of the In-48Sn solder joint and the surface finish of the BGA substrate after bonding. The In-48Sn/E-NG and In-48Sn/ENIG solder joints showed the highest and lowest shear forces among the four different solder joints, respectively. The shear force of the In-48Sn/I-Ag solder joint was higher than that of the In-48Sn/OSP solder joint.

Figure 3 shows the relationship between the displacement at break of the In-48Sn solder joint and the surface finish of the BGA substrate after bonding. The values for the I-Ag and OSP surface finishes, forming the Cu-based IMC at the interface, were higher than those for the E-NG and ENIG surface finishes. In particular, the In-48Sn/I-Ag solder joint showed the highest displacement at break among the four different solder joints.

Figure 4 shows the relationship between the work of break of the In-48Sn solder joint and the surface finish of the BGA substrate after bonding. The work of break strongly depended on the displacement at break, rather than the shear force. The In-48Sn/I-Ag and In-48Sn/ENIG solder joints exhibited the highest and lowest work of break among the four different solder joints.

The crack propagated along the weakest layer or interface, which determined the mechanical integrity of the solder joint. Therefore, failure mode has been an important acceptance criterion of the ball shear test. Figure 5 shows the fracture surface of the fractured samples after the shear tests for the In-48Sn solder joints in terms of the surface finish.

There were many $AuIn_2$ intermetallics and dimples induced by the $AuIn_2$ IMC on the fracture surface in the In-48Sn/E-NG solder joint. The $AuIn_2$ IMC improved the shear force of the In-48Sn solder joint, but degraded the displacement at break and work of break. Therefore, it was suggested that the $AuIn_2$ IMC significantly strengthened the solder joint, while the cracks easily propagated along the IMC, because of its brittleness.

The fracture surfaces of the In-48Sn/ENIG solder joints showed a mixed mode containing both ductile and brittle fractures. Many $Ni_3(Sn,In)_4$ intermetallics were observed on the fracture surface in the In-48Sn/ENIG solder joint. Therefore, it is normal that the formation of the continuous $Ni_3(Sn,In)_4$ IMC layer degraded the mechanical integrity of the solder joint, because of its brittleness. These results showed that the ENIG surface finish was unsuitable for the In-48Sn solder.

The fracture of the In-48Sn/OSP solder joints propagated along the bulk solder. The fracture surface showed typical ductile fracture characteristics. Therefore, the mechanical properties of the solder joints were determined by those of the bulk solder. During bonding, Cu was dissolved in the molten solder and the $Cu_6(Sn,In)_5$ IMC spalled off to the solder, as stated above. Therefore, the solder/OSP solder joints showed high shear properties, due to the solid-solution hardening and dispersion hardening.

The fracture characteristics of the In-Sn/I-Ag solder joints were similar to those of the In-Sn/OSP solder joints. Moreover, the shear properties of the In-Sn/I-Ag solder joints were higher than those of the In-Sn/OSP solder joints. Ag in the substrate reacted with the molten solder rapidly, and many intermetallics, spalling off to the solder, were observed in the bulk solder, as stated above. Ag and its reaction products strengthened the bulk solder better than Cu and its reaction products. Therefore, it is normal that the solder joint showed excellent mechanical properties, due to the solid-solution hardening and dispersion hardening. The I-Ag surface finish was the most suitable metallization for improving the mechanical reliability of the In-48Sn solder among the four different surface finishes tested.

## 4. CONCLUSIONS

Interfacial reactions and shear properties of the In-48Sn solder joints were investigated with four different surface finishes of the BGA substrate after bonding, to determine the optimum surface finish of the substrate for the application of the In-48Sn solder in the MOEMS package. The results are summarized as follows.

The interfacial reaction between the solder and E-NG substrate formed a continuous facetted $AuIn_2$ IMC layer during bonding. The solder/E-NG solder joint exhibited the highest shear force, while the displacement at break and work of break was low, due to the brittleness of $AuIn_2$ IMC.

During bonding, continuous $Ni_3(Sn,In)_4$ and $Ni_3P$ IMC layers were formed at the interface between the solder and ENIG substrate. The brittleness of the $Ni_3(Sn,In)_4$ IMC layer weakened the mechanical integrity of the solder joints.

$Cu_6(Sn,In)_5$ IMC, having two different morphologies, was observed after bonding. The dissolution of Cu in the molten solder and the spallation of the $Cu_6(Sn,In)_5$ IMC improved the shear properties of the solder joints.

$Cu_6(Sn,In)_5$ and $Cu(In,Sn)_2$ IMCs, with a minor Ag element, were formed at the solder/I-Ag interface during bonding. The In-48Sn/I-Ag solder joint showed the best shear properties among the four solder joints after bonding, because of the dissolution of Ag and Cu and the IMCs dispersed in the bulk solder.

## ACKNOWLEDGEMENTS

This work was supported by grant No. RTI04-03-04 from the Regional Technology Innovation Program of the Ministry of Commerce, Industry and Energy (MOCIE).

## REFERENCES

[1] Z.F. Wang, W. Cao and Z. Lu, "MOEMS: Packaging and Testing," *Microsyst Technol*, vol.12, pp. 52-58, 2005.
[2] J.M. Koo, Y.H. Lee, S.K. Kim, M.Y. Jeong and S.B. Jung, "Mechanical and Electrical Properties of Sn-3.5Ag Solder/Cu BGA Packages during Multiple Reflows," *Key Eng Mat*, vol.297-300, pp. 801-806, 2005.
[3] A.R. Mickelson, N.R. Basavanhally and Y.C. Lee, *Optoelectronic Packaging*, John Wiley & Sons, NY, pp. 209-225, 1997.
[4] J.M. Koo and S.B. Jung, "Reliability of In-48Sn solder/Au/Ni/Cu BGA Packages during Reflow Process," *J Electron Mater*, vol.34, pp. 1565-1572, 2005.
[5] J.W. Morris, Jr., J.L. Freer Goldstein and Z. Mei, "Microstructure and Mechanical Properties of Sn-In and Sn-Bi Solders," *JOM*, vol.45, pp. 25-27, 1993.
[6] J.M. Koo and S.B. Jung, "Effect of Substrate Metallization on Mechanical Properties of Sn–3.5Ag BGA Solder Joints with Multiple Reflows," *Microelectronic Eng*, vol.82, pp. 569-574, 2005.
[7] M. He, Z. Chen and G. Qi, "Solid State Interfacial Reaction of Sn–37Pb and Sn–3.5Ag Solders with Ni–P Under Bump Metallization," *Acta Mater*, vol.52, pp. 2047-2056, 2004.
[8] T.H. Chuang, C.L. Yu, S.Y. Chang and S.S. Wang, "Phase Identification and Growth Kinetics of the Intermetallic Compounds Formed during In-49Sn/Cu Soldering Reactions," *J Electron Mater*, vol.31, pp. 640-645, 2002.
[9] C.E. Ho, Y.M. Chen and C.R. Kao, "Reaction Kinetics of Solder-Balls with Pads in BGA Packages during Reflow Soldering," *J Electron Mater*, vol.28, pp.1231-1237, 1999.
[10] N.C. Lee, "Getting Ready for Lead-Free Solders," *Surf Mt Tech*, vol.9, pp. 65-69, 1997.
[11] P.G. Shewmon, *Diffusion in Solids*, Minerals, Metals & Materials Society, Warrendale, pp. 189-222, 1989.
[12] J.M. Koo, J.W. Yoon and S.B. Jung, "Interfacial Reactions between In-48Sn Solder and Electroless Nickel/Immersion Gold Substrate during Reflow Process," *Surf Interface Anal*, vol.38, In Press, 2006.
[13] D.G. Kim and S.B. Jung, "Interfacial Reactions and Growth kinetics for Intermetallic Compound Layer between In-48Sn Solder and Bare Cu Substrate," *J Alloy Compd*, vol. 386, pp. 151-156, 2005.
[14] S. Sommadossi, W. Gust and E.J. Mittemeijer, "Characterization of the Reaction Process in Diffusion-Soldered Cu/In-48 at.% Sn/Cu Joints," *Mater Chem Phys*, vol.77, pp. 924-929, 2002.
[15] T. Laurila, V. Vuorinen, J.K. Kivilahti, "Interfacial Reactions between Lead-Free Solders and Common Base Materials," *Mater Sci Eng R*, vol.49, pp.1-60, 2005.